\documentclass{aipproc}
\layoutstyle{8x11single}
\begin{document}

\title{On Counterfactuals and Contextuality}

\author{Karl Svozil}{address={Institute of Theoretical Physics, Vienna
    University of Technology, Wiedner Hauptstra\ss e 8-10/136, A-1040
    Vienna, Austria},email={svozil@tuwien.ac.at}}


\begin{abstract}
Counterfactual reasoning and contextuality is defined and critically evaluated
with regard to its nonempirical content.
To this end, a uniqueness property of states, explosion views and link observables are introduced.
If only a single context associated with a particular maximum set of observables can be operationalized,
then a context translation principle resolves measurements of different contexts.
\end{abstract}
\maketitle

\section{Counterfactuals}

With the rise of quantum mechanics
\cite{schrodinger,reich-44,jammer:66,jammer1}
physics proper entered
an ancient and sometimes fierce debate in theology and philosophy:
the controversy between realism versus idealism.
Whereas realism has been subsumed by the proposition that \cite{stace}
{\em ``some entities sometimes exist without being experienced by any finite mind,''}
idealism put forward that
{\em ``we have not the faintest reason for believing in the existence of
unexperienced entities. [[Realism]] has been adopted
$\ldots$
solely because it simplifies our view of the universe.''}
And whereas these issues can be considered nonoperational and thus metaphysical or even ideological,
it is also true that they have inspired a great number of minds,
to the effect of stimulating new approaches to quantum mechanics,
revealing many theoretical details, quantum phenomena and quantum technologies.

The Kochen-Specker theorem \cite{kochen1}, for example, was motivated from the onset by scholasticism,
as in an early programmatic article \cite{specker-60}
Ernst Specker related the discussion on the foundations of quantum mechanics
to scholastic speculations  about the existence of {\em infuturabilities.}
The scholastic issue was whether or not the omniscience (comprehensive knowledge) of God extends to
what nowadays are called {\em counterfactuals.}
And if so, can all events be pasted together to form a consistent whole?

Informally, counterfactuals will be defined as follows.
By counterfactual events we mean
{\em events which would have occurred if something had happened which did not happen.}
The associated counterfactual proposition is henceforth regarded ``true''
if it states the occurrence of an event which would have occurred if something had happened which did not happen.

Classically, at least in principle,
it makes no difference whether or not a particular observable is measured.
It is assumed to possess a definite value, irrespective of any measurement.
Thus classical counterfactuals do not present a conceptual challenge.

Quantum mechanically, the situation appears to be very different,
and the use of counterfactuals for quantized systems is problematic and unresolved.
Let us briefly mention two novel, nonrealistic features of quantum mechanics which
challenge the sort of realism suggested by classical physics.
Note, however, that also in classical times a certain uneasiness with the prevailing realistic perception
remained; e.g., in Hertz's perception of the formalisms of classical mechanics Ref. \cite{hertz-94}.

{\em Complementarity} and the {\em uncertainty principle} limit the precision of co-measurements
of certain entities.
However, nondistributive propositional structures characteristic for complementarity
not necessarily imply total abandonment of nonclassicality; e.g.,
in automaton logic \cite[Chapter 10]{svozil-ql} or generalized urn models
\cite{wright,svozil-2001-eua}.

{\em Value indefiniteness} manifests itself through the scarcity of two-valued, nondispersive
states or probability measures for particular logics or propositional structures occurring in quantum mechanics.
There are ``not enough'' two-valued states to allow a faithful embedding into a Boolean algebra.
Such two-valued states can be logically interpreted as  truth assignments.
Quantum logics \cite{pulmannova-91,svozil-ql} has developed and characterised classes of ``scarcities,''
ranging from nonunital to  nonseparable set of two-valued states to the nonexistence
of two-valued states.
The Kochen-Specker theorem \cite{kochen1}
(see also
\cite{specker-60,kamber64,kamber65,ZirlSchl-65,bell-66,Alda,Alda2,peres,mermin-93,svozil-tkadlec,tkadlec-00})
is a finitistic argument against the existence of any
consistent global truth assignment for quantized systems
associated with Hilbert spaces of dimension higher that two.

The question remains whether the set of
probability measures increases or decreases as the set of two-valued measures increases or decreases.
In classical physics and even for nondistributive logics with a separating set of two-valued states,
the answer appears to be straightforward:
By the Minkowsky-Weyl representation theorem \cite[p.29]{ziegler},
the set of probability measures is the convex hull of the set of two-valued states representable by
vertices of the associated polytope.
Due to convexity, the set of probability measures can only increase as the number of vertices increases.
But more general, e.g., quantum, probabilities are based on different assumptions.

\section{Uniqueness property of states}

In what follows the possibility of testing certain assumptions related to counterfactuals,
in particular contextuality, will be reviewed.
For the sake of the argument, certain properties of states will be defined,
and the operationalization of explosion views of theoretical arguments will be discussed.

\subsection{Definition}

A multiquantum state will be said to satisfy the {\em uniqueness property} if
knowledge of a property of one quantum entails the certainty
that, if this property
were measured on the other quantum (or quanta) as well, the outcome of the measurement would be
a unique function of the outcome of the measurement performed.
This uniqueness property could be experimentally tested by performing the associated experiments
for co-measurable observables of different quanta.

Einstein, Podolsky and Rosen \cite{epr} have proposed a way of counterfactual interference
of particle properties by using certain entangled two-quantum states with the uniqueness property.
Suppose one is willing to accept counterfactuals.
Then
one may pretend to obtain knowledge of
noncommeasurable observables referring to a single quantum
by measurement of one observable per quantum
in a multiquantum state satisfying the uniqueness property at a time,
and by subsequent counterfactual inference.
This assumption (to be falsified) is also used in the Kochen-Specker {\em reductio ad absurdum} proof.

This method could in principle be generalized to $n$ quanta, such that
the measurement of the state of one quantum
in an $n$-quantum system fixes
the states of all the others as well.

\subsection{Example states}

Apparently, the uniqueness property can be satisfied for multipartite states
for which the number of terms contributing to
the coherent sum of the joint amplitudes does not exceed the dimension of the
single particle Hilbert space.
Indeed, for nontrivial configurations, the number of terms should be {\em identical}
to the single particle Hilbert space dimension.
For if the number of terms exceeds the dimension, then
at least for one quantum there are more than one possibilities of counterfactual existence.
Consider, for example, three spin one quanta.
Their only singlet state is (see also \cite{kok-02})
\begin{equation}
\vert \Psi_3 \rangle
= {1\over \sqrt{6}}(
\vert - + 0\rangle
-
\vert - 0 +\rangle
+
\vert + 0 - \rangle
-
\vert + - 0\rangle
+
\vert 0 - + \rangle
-
\vert 0 + - \rangle
).
\label{2004-qnc-e1}
\end{equation}
Now, suppose the outcome of a spin measurement on the first quantum is ``$-$;''
serving as a filter and reducing $\vert \Psi_3 \rangle$
to the first two terms
$$({1/ \sqrt{2}})(
\vert - + 0\rangle
-
\vert - 0 +\rangle
)$$
in the coherent sum of Eq.~(\ref{2004-qnc-e1}).
Thereby, two possibilities ``$0$''
and ``$+$'' remain for the state of every one of the other quanta.
This ambiguity in the counterfactual argument results in nonuniqueness.

With regards to uniqueness, the situation gets worse for singlet states of four or more spin one quanta;
e.g., the three states $\vert \Psi_4^1\rangle , \vert \Psi_4^2\rangle , \vert \Psi_4^3\rangle$
of four spin one quanta \cite{schimpf-svozil}
\begin{eqnarray}
\vert \Psi_4^1\rangle  &=&
\frac{1}{\sqrt{5}}
\big[
\frac{2}{3}
\vert0,0,0,0\rangle
+
\vert -1,-1,1,1\rangle  +\vert 1,1,-1,-1\rangle
\nonumber \\
&&\quad
- \frac{1}{2}
\big(
\vert-1,0,0,1\rangle
+\vert0,-1,0,1\rangle
+\vert-1,0,1,0\rangle
+\vert0,-1,1,0\rangle \nonumber   \\
&&\quad
\qquad
+\vert0,1,-1,0\rangle
+\vert 1,0,-1,0\rangle
+\vert0,1,0,-1\rangle
+\vert1,0,0,-1\rangle
\big)\nonumber \\
&&\quad
+\frac{1}{3}
\big(
\vert0,0,-1,1\rangle
+\vert -1,1,0,0\rangle
+\vert1,-1,0,0\rangle
+\vert0,0,1,-1\rangle
\big)\nonumber \\
&&\quad
+
\frac{1}{6}
\big(
\vert-1,1,-1,1\rangle
+\vert1,-1,-1,1\rangle
+\vert-1,1,1,-1\rangle
+\vert1,-1,1,-1\rangle
\big)
\big] ,
 \\
\vert \Psi_4^2\rangle  &=&\frac{1}{2{\sqrt{3}}}
\big(
\vert -1,0,0,1\rangle
- \vert 0,-1,0,1\rangle
- \vert 0,1,0,-1\rangle
+\vert 1,0,0,-1\rangle  \nonumber \\
&&\quad
- \vert -1,0,1,0\rangle
+\vert 0,-1,1,0\rangle
+\vert 0,1,-1,0\rangle
- \vert 1,0,-1,0\rangle \nonumber    \\
&&\quad
+\vert -1,1,1,-1\rangle
- \vert 1,-1,1,-1\rangle
- \vert -1,1,-1,1\rangle
+\vert 1,-1,-1,1\rangle
\big) ,
  \\
  \vert \Psi_4^3\rangle &=& \frac{1}{3}
\big(
\vert 0,0,0,0\rangle
-
\vert 0,0,-1,1\rangle
-
\vert -1,1,0,0\rangle
-
\vert 1,-1,0,0\rangle
-
\vert 0,0,1,-1\rangle
\nonumber  \\
&& \quad
+
\vert 1,-1,1,-1\rangle
+
\vert -1,1,-1,1\rangle
+
\vert 1,-1,-1,1\rangle
+
\vert -1,1,1,-1\rangle
\big) ,
\end{eqnarray}
do not hold a uniqueness property, as can be readily verified with the same argument as above.

For singlet states of quanta of three dimensions, the uniqueness limit is reached for two particles
$$
\vert \Psi_2 \rangle
= ({1/ \sqrt{3}})(
\vert + -\rangle
+
\vert - +\rangle
-
\vert 0 0\rangle
).$$
For less terms the configuration may be effectively lower dimensional.

As already stated, for measuring entangled particles at different contexts, the uniqueness property
must hold for {\em every} such context.
This additional assumption is satisfied for singlet states,
which are form invariant under identical unitary transformations of the single quantum Hilbert spaces.
This is not true for the Greenberger, Horne and Zeilinger three spin one half quanta state
in the form proposed by Mermin \cite{mermin1}
$$
\vert \Psi_{GHZM} \rangle
= ({1/ \sqrt{2}})(
\vert z+ z+ z+\rangle
+
\vert z- z- z-\rangle
).$$
Here $z\pm$ stands for the outcome $\pm$ of a spin measurement  measured along the $z$-axis.
A careful calculation (e.g., Eq.~(3) of Ref.~\cite{krenn1}) shows that
$\vert \Psi_{GHZM}\rangle $
satisfies the uniqueness property only along a single direction, the $z$-axis, of spin state measurements;
otherwise $\vert \Psi_{GHZM}\rangle $ contains eight summands.

Whether or not nontrivial states (in the sense mentioned above) exist which
satisfy the uniqueness property in ``sufficiently many'' spin state measurement directions
to make them useful for conterfactual reasoning
remains an open question
Alas, the lack of uniqueness  may be the reason why inconsistencies such as the ones derived in the
Kochen-Specker type proofs cannot be directly operationalized;
such as in the ``explosion type'' nonclassical setups discussed below.

\section{Explosion views}

If the quantum state also satisfies the uniqueness property when transformed to different,
complementary measurements, then
different,  complementary, observables on other quanta could be measured, for which a similar
uniqueness property holds.
In that way, one may pretend to obtain knowledge of
all these noncommeasurable observables referring to a single quantum
by measurement of one context per quantum at a time, and by subsequent counterfactual inference.
Of course, only one of these properties would actually be obtained by direct measurement on the
quantum; all the other properties are merely counterfactually inferred.
In principle, this method can be applied to an arbitrary number of contexts and quanta
as long as the uniqueness property holds.
This kind of setups will be referred to as of the {\em ``explosion view''} type.

The advantage of explosion views
is that they do not require different terms which refer to different detector settings and thus to measurements
performed at different times.
On the contrary, all  standard Bell-Clauser-Horne-Shimony-Halt type
or Greenberger-Horne-Zeilinger type measurements \cite{panbdwz} known today
involve the summation or consideration of terms which are not co-measurable.
The inevitable time delay of consecutive measurements of the relevant noncomeasurable terms
makes these arguments vulnerable to a
critique put forward by Hess and Philipp \cite{Hess&Philipp2002}.
Furthermore, explosion views offer
the possibility to directly measure contextuality.

In this counterfactual sense, the measurement of observables associated with arbitrary operators
becomes feasible, since formally, any matrix $A$ can be decomposed into two
self-adjoint components
$A_1, A_2$ as follows:
\begin{equation}
\begin{array}{lllllll}
A&=&A_1+iA_2 \\
A_1&=&{1\over 2}(A+A^\dagger) =:\Re A,\;   \;
A_2=-{i\over 2}(A-A^\dagger)=:\Im A.
\end{array}  \label{e-decom1}
\end{equation}
By assuming the uniqueness property,  $A_1, A_2$ can be measured along two different entangled
particles, respectively, and subsequently counterfactually ``completed.''

\section{Contextuality}

There exist various notions of contextuality.
In what follows, the term contextuality will be used as envisioned by Bell and Redhead
\cite{bell-66,hey-red,redhead}.
Already
Bohr~\cite{bohr-1949} mentioned
{\em ``the impossibility of any sharp separation
between the behavior of atomic objects and the interaction with the measuring instruments which serve to define
the conditions under which the phenomena appear.''}
Bell (Ref.~\cite{bell-66}, Sec.~5) stated that the {\em ``$\ldots$
result of an observation may reasonably depend
not only on the state of the system  $\ldots$
but also on the complete disposition  of the apparatus.''}
That is, the outcome of the measurement of an observable  $A$
might depend on which other observables
from systems of maximal observables
are measured alongside with $A$.

This concept was mainly introduced to maintain a certain amount of realism
in view of the challenges of the theorems by Gleason \cite{Gleason} and Kochen and Specker;
i.e., the ``scarcity'' of two valued states mentioned above.
(Other attempts towards this goal have assumed nonconstructive measure theory
utilizing paradoxical set decomposition
\cite{pitowsky-82,pitowsky-83},
or abandoned the continuity of Hilbert space \cite{meyer:99,havlicek-2000}.)
This section presents a critical evaluation of its empirical content.

\subsection{Context and link observables}

A {\em context} can formally be defined as a single (nondegenerate) ``maximal'' self-adjoint operator  ${ C}$.
It has a spectral
decomposition into some complete set of orthogonal projectors ${ E}_i$
which correspond to propositions in
the usual Von Neumann-Birkhoff type sense \cite{birkhoff-36,v-neumann-49}.
That is, ${ C}=\sum_{i=1}^n e_i { E}_i$
with mutually different $e_i$ and some orthonormal basis $\{{ E}_i{ H} \mid i=1,\ldots n\}$ of
$n$-dimensional Hilbert space ${ H}$.
In $n$ dimensions, contexts can be viewed as $n$-pods spanned by the $n$ orthogonal projectors
${ E}_1,
{ E}_2, \cdots,{ E}_n$.

In the finite subalgebras considered, an observable belonging to two or more contexts is called {\em link observable}.

Contexts can thus be depicted by Greechie (orthogonality) diagrams
\cite{greechie:71},  consisting of {\em points} which
symbolize observables (representable by the spans of vectors
in $n$-dimensional Hilbert space).
Any $n$ points belonging to a context; i.e., to a maximal set of commeasurable observables
(representable as some orthonormal basis of  $n$-dimensional Hilbert space),
are connected by {\em smooth curves}.
Two smooth curves may be crossing in  common {\em link observables}.
In three dimensions, smooth curves and the associated points stand for tripods.
Still another compact representation is in terms of Tkadlec diagrams,
where points represent complete tripods and smooth curves represent
single legs interconnecting them.
In quantum logic \cite{pulmannova-91,svozil-ql,kalmbach-83},
contexts are often referred to as {\em subalgebras} or {\em blocks.}

\subsection{Experimental falsification of contextuality in simple configurations: Two contexts in three dimensions}

In two dimensional Hilbert space, contextuality does not exist,
since every context is fixed by the assumption of one property.
The entire context is just this property, together with its negation,
which corresponds to the orthogonal ray (which spans a one dimensional subspace)
or projection associated with the ray corresponding to the property.

\begin{figure}
\begin{tabular}{ccccc}
\unitlength 0.70mm
\linethickness{0.4pt}
\begin{picture}(40.00,49.67)
\put(15.00,45.00){\line(0,-1){30.00}}
\put(15.00,15.00){\line(-1,-1){15.00}}
\put(15.00,15.00){\line(1,0){25.00}}
\put(15.00,15.00){\line(3,-4){11.00}}
\put(15.00,15.00){\line(5,3){16.67}}
\put(3.33,-1.67){\makebox(0,0)[cc]{$B$}}
\put(30.00,0.00){\makebox(0,0)[cc]{$D$}}
\put(40.00,11.33){\makebox(0,0)[cc]{$C$}}
\put(35.00,23.67){\makebox(0,0)[cc]{$K$}}
\put(19.33,49.67){\makebox(0,0)[cc]{$A$}}
\put(20.00,6.67){\vector(2,1){0.2}}
\bezier{60}(7.67,6.67)(14.33,2.67)(20.00,6.67)
\put(30.00,23.00){\vector(-1,2){0.2}}
\bezier{36}(29.67,16.00)(32.33,19.67)(30.00,23.00)
\put(13.33,1.00){\makebox(0,0)[cc]{$\varphi$}}
\put(40.00,18.67){\makebox(0,0)[rc]{$\varphi$}}
\end{picture}
&&
\unitlength 0.80mm
\linethickness{0.4pt}
\begin{picture}(61.33,36.00)
\multiput(0.33,35.00)(0.36,-0.12){84}{\line(1,0){0.36}}
\multiput(30.33,25.00)(0.36,0.12){84}{\line(1,0){0.36}}
\put(30.33,25.00){\circle{2.00}}
\put(45.33,30.00){\circle{2.00}}
\put(60.33,35.00){\circle{2.00}}
\put(0.33,35.00){\circle{2.00}}
\put(15.33,30.00){\circle{2.00}}
\put(60.33,31.00){\makebox(0,0)[cc]{$K$}}
\put(45.33,26.00){\makebox(0,0)[cc]{$D$}}
\put(30.33,30.00){\makebox(0,0)[cc]{$A$}}
\put(15.33,26.00){\makebox(0,0)[cc]{$C$}}
\put(0.33,31.00){\makebox(0,0)[cc]{$B$}}
\bezier{24}(0.00,20.00)(0.00,17.33)(3.00,17.33)
\bezier{28}(3.00,17.33)(10.00,17.00)(10.00,17.00)
\bezier{32}(10.00,17.00)(15.00,16.00)(15.00,13.33)
\bezier{24}(30.00,20.00)(30.00,17.33)(27.00,17.33)
\bezier{28}(27.00,17.33)(20.00,17.00)(20.00,17.00)
\bezier{32}(20.00,17.00)(15.00,16.00)(15.00,13.33)
\put(15.00,5.33){\makebox(0,0)[cc]{$\{B,C,A\}$}}
\bezier{24}(60.00,20.00)(60.00,17.33)(57.00,17.33)
\bezier{28}(57.00,17.33)(50.00,17.00)(50.00,17.00)
\bezier{32}(50.00,17.00)(45.00,16.00)(45.00,13.33)
\bezier{24}(30.00,20.00)(30.00,17.33)(33.00,17.33)
\bezier{28}(33.00,17.33)(40.00,17.00)(40.00,17.00)
\bezier{32}(40.00,17.00)(45.00,16.00)(45.00,13.33)
\put(45.00,5.33){\makebox(0,0)[cc]{$\{A,D,K\}$}}
\end{picture}
&&
\unitlength 0.80mm
\linethickness{0.4pt}
\begin{picture}(51.37,10.00)
\put(0.00,25.00){\circle{2.75}}
\put(30.00,25.00){\circle{2.75}}
\put(1.33,25.00){\line(1,0){27.33}}
\put(15.00,30.00){\makebox(0,0)[cc]{$A$}}
\put(30.00,15.00){\makebox(0,0)[cc]{$\{A,D,K\}$}}
\put(0.00,15.00){\makebox(0,0)[cc]{$\{B,C,A\}$}}
\end{picture}
\\
a)&&b)&&c)\\
\end{tabular}
\caption{Three equivalent representations of the same geometric configuration:
a) Two tripods with a common leg;
b) Greechie (orthogonality) diagram: points stand for individual basis vectors, and
orthogonal tripods are drawn as smooth curves;
c) Tkadlec diagram: points represent complete tripods and smooth curves represent
single legs interconnecting them.
\label{2004-qnc-f1}}
\end{figure}
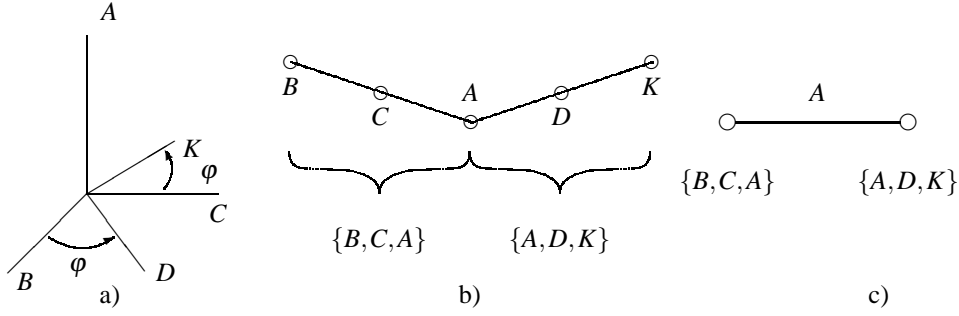

The simplest nontrivial configuration of contexts exist in three dimensional Hilbert space.
Consider an arrangement
of five observables $A,B,C,D,K$ with two systems
of operators
$\{A,B,C\}$
and
$\{D,K,A\}$
called {\em contexts},
which are interconnected by $A$.
With a context, the operators commute and the associated observables are commeasurable.
For two different contexts, operators outside the link operators do not commute.
$A$ will be called a {\em link observable}.
This propositional structure can be represented in three dimensional Hilbert space
by two tripods with a single common leg.
Fig.~\ref{2004-qnc-f1} depicts this configuration in three dimensional real vector space,
as well as in the associated Greechie and Tkadlec diagrams.
The operators  $B,C,A$ and $D,K,A$ can be identified with the projectors corresponding
to the two bases
\begin{equation}
\begin{array}{lcl}
B_{B-C-A}&=&
\{
(1,0,0)^T,
(0,1,0)^T,
(0,0,1)^T
\}
,
\\
B_{D-K-A}&=&
\{
(\cos \varphi , \sin \varphi ,0)^T,
(-\sin \varphi ,\cos \varphi , 0)^T,
(0,0,1)^T
\},
\end{array}
\label{e-vaxjo1}
\end{equation}
(the superscript ``$T$'' indicates transposition).
Their matrix representation is the  dyadic product of every vector with itself.

Physically, the union of contexts $\{B,C,A\}$ and $\{D,K,A\}$ interlinked along $A$ does not have any direct
operational meaning; only a single context can be measured along a single quantum at a time;
the other being irretrievably lost if no reconstruction of the original state is possible.
Thus, in a direct way, testing the value of observable $A$ against different
contexts $\{B,C,A\}$ and $\{D,K,A\}$ is metaphysical.

It is, however, possible to counterfactually retrieve information
about the two different contexts of a single quantum indirectly
by considering a singlet state
$
\vert \Psi_2 \rangle
= ({1/ \sqrt{3}})(
\vert + -\rangle
+
\vert - +\rangle
-
\vert 0 0\rangle
)$
via the ``explosion view'' Einsten-Podolsky-Rosen type of argument discussed above.
Since the state is form invariant with respect to variations of the measurement angle
and at the same time satisfies the uniqueness property,
one may retrieve the first context
$\{B,C,A\}$ from the first quantum
and the second context $\{D,K,A\}$ from the second quantum.
(This is a standard procedure in Bell type arguments with two spin one-half quanta.)

In this indirect, counterfactual sense, contextuality becomes measurable.
From an experimental point of view, this amounts to performing two tasks.
\begin{itemize}
\item[(i)]
In the preparation stage,
a singlet state of two spin one quanta must be realized.
This has become feasible recently by
engineering entangled
states in any arbitrary dimensional Hilbert space
\cite{mvwz-2001,vwz-2002,gisin-2002-d,tdtm-2003}
(see also generalized beam splitter setups  \cite{rzbb,reck-94,zukowski-97,svozil-2004-analog} for proofs of principle).
\item[(ii)]
In the analyzing stage,
the context structure  $\{B,C,A\}$ and $\{D,K,A\}$ interlinked at $A$ must be realized.
\end{itemize}
Let the matrix $[{ v}^T{ v}]$
stand for the dyadic product
of the vector ${ v}$ with itself.
The operators
associated with the geometrical configuration enumerated in Eq.~(\ref{e-vaxjo1})
depicted in Fig.~\ref{2004-qnc-f1}a) are given by
\begin{equation}
\begin{array}{lcl}
C_{B-C-A}&=& \sum_{i=1,2,3} e_i [B_{B-C-A,i}^T B_{B-C-A,i}]=
\left(
\begin{array}{ccc}
e_1&0&0\\
0&e_2&0\\
0&0&e_3\\
\end{array}
\right)
,
\\
C_{D-K-A}
&=& \sum_{i=1,2,3} e_i' [B_{D-K-A,i}^T B_{D-K-A,i}]=
\left(
\begin{array}{ccc}
e_1' \cos^2 \varphi + e_2'\sin^2 \varphi&(e_1'-e_2')\sin \varphi \cos \varphi &0\\
(e_1'-e_2')\sin \varphi \cos \varphi &e_2' \cos^2 \varphi + e_1'\sin^2 \varphi &0\\
0&0&e_3'\\
\end{array}
\right).
\end{array}
\label{e-vaxjo2}
\end{equation}
As mentioned before,
$e_1,e_2,e_3$
as well as
$e_1',e_2',e_3'$
must be mutually distinct.

Contextuality would imply that the outcome of a measurement of $A$ depends on which other operators
are measured alongside with it.
If contextuality were taken seriously for the two context configuration
discussed above, the measured value of $A$ would be different for the first and for the second quantum
in the entangled two particle singlet state $\vert \Psi_3 \rangle$.
Note that the context structure discussed here not necessarily implies that
$A$ corresponds to the joint event associated with ``$00$,''
as the labelling applied to the state preparation has nothing to do with the
logico-algebraic propositional structure realized by the contexts.

Despite the necessity to falsify contextuality for the interlinked two-context structure experimentally,
there can hardly be any doubt about the outcome of the verdict against contextuality.
This is due to the symmetry of the prepared singlet state,
which is an expression of the conservation laws of quantities such as angular momentum.
A conceivable option to save contextuality would be to assume that only in the nununique state cases
contextuality unfolds.
However, this would make contextuality a metaphysical property which cannot be measured at all
and which has no physical meaning whatsoever.

\subsection{Three contexts in three dimensions}

\begin{figure}
\begin{tabular}{ccccc}
\unitlength 0.70mm
\linethickness{0.4pt}
\begin{picture}(91.34,36.00)
\multiput(0.33,35.00)(0.36,-0.12){84}{\line(1,0){0.36}}
\put(30.33,25.00){\circle{2.00}}
\put(45.33,25.00){\circle{2.00}}
\put(0.33,35.00){\circle{2.00}}
\put(15.33,30.00){\circle{2.00}}
\bezier{24}(0.00,20.00)(0.00,17.33)(3.00,17.33)
\bezier{28}(3.00,17.33)(10.00,17.00)(10.00,17.00)
\bezier{32}(10.00,17.00)(15.00,16.00)(15.00,13.33)
\bezier{24}(30.00,20.00)(30.00,17.33)(27.00,17.33)
\bezier{28}(27.00,17.33)(20.00,17.00)(20.00,17.00)
\bezier{32}(20.00,17.00)(15.00,16.00)(15.00,13.33)
\put(15.00,5.33){\makebox(0,0)[cc]{$\{B,C,A\}$}}
\bezier{24}(60.67,20.00)(60.67,17.33)(57.67,17.33)
\bezier{28}(57.00,17.33)(50.00,17.00)(50.00,17.00)
\bezier{32}(50.00,17.00)(45.00,16.00)(45.00,13.33)
\bezier{24}(30.00,20.00)(30.00,17.33)(33.00,17.33)
\bezier{28}(33.00,17.33)(40.00,17.00)(40.00,17.00)
\bezier{32}(40.00,17.00)(45.00,16.00)(45.00,13.33)
\put(45.00,5.33){\makebox(0,0)[cc]{$\{A,D,K\}$}}
\put(30.33,25.00){\line(1,0){30.00}}
\multiput(90.34,35.00)(-0.36,-0.12){84}{\line(-1,0){0.36}}
\put(60.34,25.00){\circle{2.00}}
\put(90.34,35.00){\circle{2.00}}
\put(75.34,30.00){\circle{2.00}}
\bezier{24}(90.67,20.00)(90.67,17.33)(87.67,17.33)
\bezier{28}(87.67,17.33)(80.67,17.00)(80.67,17.00)
\bezier{32}(80.67,17.00)(75.67,16.00)(75.67,13.33)
\bezier{24}(60.67,20.00)(60.67,17.33)(63.67,17.33)
\bezier{28}(63.67,17.33)(70.67,17.00)(70.67,17.00)
\bezier{32}(70.67,17.00)(75.67,16.00)(75.67,13.33)
\put(75.67,5.33){\makebox(0,0)[cc]{$\{K,L,M\}$}}
\end{picture}
&  &
\unitlength 0.80mm
\linethickness{0.4pt}
\begin{picture}(40.52,42.02)
\put(0.00,9.98){\line(1,0){40.02}}
\put(40.02,9.98){\line(-4,5){20.20}}
\put(19.82,35.23){\line(-4,-5){20.20}}
\put(0.00,4.99){\makebox(0,0)[cc]{$A$}}
\put(19.96,4.99){\makebox(0,0)[cc]{$B$}}
\put(40.02,4.99){\makebox(0,0)[cc]{$C$}}
\put(35.03,24.95){\makebox(0,0)[cc]{$L$}}
\put(19.96,42.02){\makebox(0,0)[cc]{$K$}}
\put(4.99,24.95){\makebox(0,0)[cc]{$D$}}
\put(19.96,9.98){\circle{1.00}}
\put(19.86,35.13){\circle{1.00}}
\put(40.02,9.98){\circle{1.00}}
\put(30.04,22.46){\circle{1.00}}
\put(9.68,22.46){\circle{1.00}}
\put(0.00,9.98){\circle{1.00}}
\end{picture}
\\
a)& &b)
\end{tabular}
\caption{Greechie diagrams of three interlinked contexts. Case b) has no realization in Hilbert space.
\label{2004-vaxjo-f2}}
\end{figure}
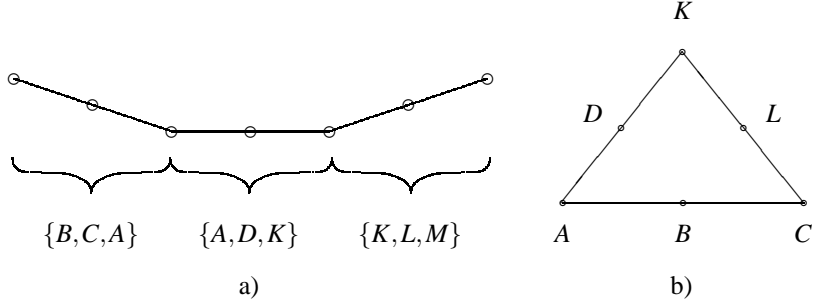

A next step further would be the logico-algebraic propositional structure
with three interlinked contexts such as
$\{A,B,C\}$,
$\{A,D,K\}$
and
$\{K,L,M\}$ interconnected at $A$ and $K$.
This configuration is depicted in Fig.~\ref{2004-vaxjo-f2}a).
Here, for the first time, the Einstein-Podolsky-Rosen ``explosion view'' type of setup
encounters the problem of nonuniqueness:
for the three quantum singlet state
$
\vert \Psi_3 \rangle
$
enumerated in
Eq.~(\ref{2004-qnc-e1})
the uniqueness property does not hold.

Note also that too tightly interlinked contexts are not realizable in Hilbert space.
The interconnected ``triangular'' system of contexts
$\{A,B,C\}$,
$\{A,D,K\}$
and
$\{K,L,C\}$  drawn in Fig.~\ref{2004-vaxjo-f2}b)
has no representation as operators in Hilbert space.
Likewise, no system of three tripods exist such that every tripods is interlinked
with the other two tripods in two different legs.

\subsection{Explosion views of more elaborate contexts in three dimensions}

The same nonuniqueness problems plaguing already the three quanta singlet state
get worse for logico-algebraic configurations with a higher number of interlinked contexts.
This makes impossible a ``direct''
counterfactual inference of the Einsten-Podolsky-Rosen type
in the case of Kochen-Specker type proofs; i.e., for
configurations supporting only nununital, nonseparating or even nonexisting sets of
two-valued probability measures corresponding to classical truth assignments.
All these nonclassical configurations lack essential classical properties.
Note that, in general, the nonexistence of any two-valued state is a very strong property requiring ``a lot more''
contexts than more subtle nonclassical features such as nununitality
or nonseparability \cite{svozil-ql}.

\begin{figure}
\unitlength 0.50mm
\linethickness{0.4pt}
\begin{picture}(190.67,109.67)
\multiput(165.67,19.67)(-0.12,0.12){167}{\line(-1,0){0.12}}
\put(145.67,39.67){\line(0,1){40.00}}
\multiput(145.67,79.67)(0.12,0.12){167}{\line(0,1){0.12}}
\multiput(165.67,99.67)(0.12,-0.12){167}{\line(0,-1){0.12}}
\put(185.67,79.67){\line(0,-1){40.00}}
\multiput(185.67,39.67)(-0.12,-0.12){167}{\line(-1,0){0.12}}
\put(185.34,59.67){\line(-1,0){39.67}}
\put(185.67,39.67){\circle{2.00}}
\put(185.67,59.67){\circle{2.00}}
\put(185.67,79.67){\circle{2.00}}
\put(145.67,39.67){\circle{2.00}}
\put(145.67,59.67){\circle{2.00}}
\put(145.67,79.67){\circle{2.00}}
\put(165.67,19.67){\circle{2.00}}
\put(165.67,99.67){\circle{2.00}}
\multiput(165.67,19.67)(-0.21,0.12){334}{\line(-1,0){0.21}}
\multiput(95.67,59.67)(0.21,0.12){334}{\line(1,0){0.21}}
\put(95.67,59.67){\circle{2.00}}
\put(95.67,74.67){\makebox(0,0)[cc]{$a_8=a_8'$}}
\put(165.67,109.67){\makebox(0,0)[cc]{$b=a_9=a_0'$}}
\put(140.67,79.67){\makebox(0,0)[cc]{$a_2'$}}
\put(140.67,59.67){\makebox(0,0)[cc]{$a_6'$}}
\put(140.67,39.67){\makebox(0,0)[cc]{$a_4'$}}
\put(190.67,39.67){\makebox(0,0)[cc]{$a_3'$}}
\put(190.67,59.67){\makebox(0,0)[cc]{$a_5'$}}
\put(190.67,80.00){\makebox(0,0)[cc]{$a_1'$}}
\put(165.67,9.67){\makebox(0,0)[cc]{$a_7'$}}
\multiput(25.00,19.67)(0.12,0.12){167}{\line(1,0){0.12}}
\put(45.00,39.67){\line(0,1){40.00}}
\multiput(45.00,79.67)(-0.12,0.12){167}{\line(0,1){0.12}}
\multiput(25.00,99.67)(-0.12,-0.12){167}{\line(0,-1){0.12}}
\put(5.00,79.67){\line(0,-1){40.00}}
\multiput(5.00,39.67)(0.12,-0.12){167}{\line(1,0){0.12}}
\put(5.33,59.67){\line(1,0){39.67}}
\put(5.00,39.67){\circle{2.00}}
\put(5.00,59.67){\circle{2.00}}
\put(5.00,79.67){\circle{2.00}}
\put(45.00,39.67){\circle{2.00}}
\put(45.00,59.67){\circle{2.00}}
\put(45.00,79.67){\circle{2.00}}
\put(25.00,19.67){\circle{2.00}}
\put(25.00,99.67){\circle{2.00}}
\multiput(25.00,19.67)(0.21,0.12){334}{\line(1,0){0.21}}
\multiput(95.00,59.67)(-0.21,0.12){334}{\line(-1,0){0.21}}
\put(25.00,109.67){\makebox(0,0)[cc]{$a=a_0=a_9'$}}
\put(50.00,79.67){\makebox(0,0)[cc]{$a_2$}}
\put(50.00,59.67){\makebox(0,0)[cc]{$a_6$}}
\put(50.00,39.67){\makebox(0,0)[cc]{$a_4$}}
\put(-0.00,39.67){\makebox(0,0)[cc]{$a_3$}}
\put(-0.00,59.67){\makebox(0,0)[cc]{$a_5$}}
\put(-0.00,80.00){\makebox(0,0)[cc]{$a_1$}}
\put(25.00,9.67){\makebox(0,0)[cc]{$a_7$}}
\put(15.00,89.67){\circle{2.00}}
\put(35.33,89.34){\circle{2.00}}
\put(25.00,59.67){\circle{2.00}}
\put(15.00,29.67){\circle{2.00}}
\put(35.00,29.67){\circle{2.00}}
\put(155.67,89.67){\circle{2.00}}
\put(176.00,89.34){\circle{2.00}}
\put(165.67,59.67){\circle{2.00}}
\put(155.67,29.67){\circle{2.00}}
\put(175.67,29.67){\circle{2.00}}
\end{picture}
\caption{\label{f-ksg3}
Greechie (orthogonality) diagram
of a Hilbert lattice
 with a nonseparating set of probability measures
 (see Graph $\Gamma_3$ of Ref.~\protect\cite{kochen1}).  \label{2004-qnc-f2}
}
\end{figure}
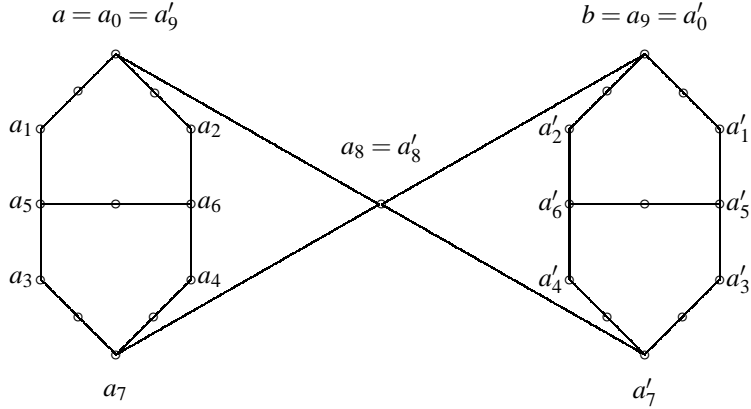

For example, the logico-algebraic structure of observables
depicted in Fig.~\ref{2004-qnc-f2}
are representable by quanta in three dimensional Hilbert space.
It has the nonclassical feature of a nonseparating set of two-valued probability measures:
For all two valued probability measures $P(x)\in\{0,1\}$, $P(a)=P(b)=1$,
there is no probability measure separating $a$ from $b$ through
$P(a)\neq P(b)$.

That is, if $P(a=a_0=a_9') = 1$ for any
two-valued probability measure $P$, then $P(a_8) =0$.
Furthermore,
$P(a_7)=0$, since by a similar argument $P(a)=1$ implies $P(a_7)=0$.
Therefore, $P(b=a_9=a_0') = 1$. Symmetry requires that the
reverse implication is also fulfilled, and therefore  $P(b) =
P(a)$ for every two-valued probability measure $P$.

An explosion requires 16 contexts and
could in principle be realized with some singlet state of 16 spin-one quanta;
a state which contains by far too many terms to satisfy the uniqueness property.
Here, nonuniquess seems to serve as a kind of {\em protection principle}
in the case of nonclassical features of two-valued probability measures,
the extreme case being the nonexistence of any such measure
(cf. Graph $\Gamma_2$ of Ref.~\protect\cite{kochen1}).

\section{Principle of context translation}

Suppose it is not unreasonable to
speculate about the following three assumptions \cite{svozil-2003-garda}:
\begin{itemize}
\item[(i)] {\em Unique context preparation:}
It is possible to encode into quanta
a certain finite amount of information by preparing them in a single context.
This amount is determined by the dimension of the associated Hilbert space.
\item[(ii)] {\em Nonexistence of different contexts:}
(Counterfactual) Elements of physical reality which go beyond that single context
do not exist.
\item[(iii)] {\em Context translation principle:}
If quanta are measured along a different context,
that context may be translated by the measurement apparatus
into the context the quanta have been originally prepared for.
The capability of the measurement apparatus to translate the context
may depend on certain parameters, such as temperature.
\end{itemize}

As strange as the first assumption may appear,
it amounts to the everyday experience
that no agent, deterministic or other,
can be prepared to render answers to every conceivable question.
A ``silly'' example for this feature would be the attempt of a person
to enter the question
{\em ``is there enough oil in the car's engine?''} at
the command prompt of a desktop computer.
Most likely, the machine would respond with some sort of
statement that this input is not recognized as a command,
executable, or batch file.
Within the standalone desktop context, the question makes no sense at all.
Yet nobody would come up with the suggestion that something strange,
bordering to the mysterious, or ``mindboggling'' is going on.
Compare also Zeilinger's {\em ``Foundational Principle''}
stating that $n$ elementary two-state systems carry $n$ bits \cite{zeil-99,zeil-bruk-02}.

When considering quantized spin systems, one often has classical angular momentum models in mind;
an observable which can be defined precisely in all directions.
Quantum mechanically, this is no longer the case.
Hence, some classical properties have to be given up.
Spin or quantized angular momentum cannot be conceive as something
being precisely defined in all directions simultaneously;
it can be only defined precisely in a singular direction.

Nevertheless, spin measurements in different directions {\em do}
give results, albeit randomized ones.
Here, the context translation principle (iii) might be assumed,
stating that any measurement apparatus capable of measuring different contexts from the one
the quantum was originally prepared for, performs some kind of translation between the contexts.
This translation may be thought of as brought about
by intrinsic microscopic processes in the measurement apparatus.
Take, as an analogy, linearly polarized light along a particular
direction, and a linear polarization measurement in a different direction.
Due to the dynamics of the oriented macromolecules of the polarization filters,
there is light leaving the measurement filter
(if it is not oriented perpendicular to the original polarization direction),
and its polarization direction is changed to the orientation of the measurement device.

Assumptions (i) and (ii) amount to a subtle form of realism,
since some ``maximal'' property---the context in which the quantum has been prepared for---is assumed to
exist even without being observed by any finite mind.
Yet other properties, associated with different contexts, are assumed to not exist at all.
The possibility to measure these nonexisting properties presents an illusion,
which is mediated by the ability of the measurement apparatus to translate
the measurement context into the prepared context.
There is no general method
to obtain knowledge of an unknown specific preparation context for individual quanta.
Neither is it possible to obtain {\it a posteriori} knowledge of such a context.

In this regard, quanta behave like little universal automata capable of storing a multitude of properties;
yet only a single property at a time.
The realm of this property is defined by the dimension of Hilbert space associated with this quantum.

\section{Summary and open questions}

This paper contains two main threads: a critical evaluation of counterfactuals, in particular contextuality,
and a discussion of a context translation principle.
Although there are strong connections,
both issues could also be perceived independently.

The experimental and theoretical status can be summarized as follows.
Measurements on the context (in)dependence
of two-context two quanta (in three dimensions per quantum) configurations
are feasible but still need to be done.

It remains an open theoretical question whether or not
nonsinglet states exist which satisfy the nuniqueness property for sufficiently many
different measurement ``directions'' or setups to allow for
``explosion views'' of more than two contexts by Einstein-Podolsky-Rosen type ``explosion view''
multipartite configurations.
There maybe some higherdimensional states which satisfy a uniqueness property in some lowerdimensional subspace.
If, as the author suspects, no such states exist,
then the case against contextuality is quite firm.
For, insofar as contextuality could be operationalized, it is likely to be falsified;
and in more nontrivial cases it could not be operationalized.
To state it pointedly, contextuality might turn out to be a {\em red herring.}

Finally, the entire issue of context translation remains speculative and
theoretically and experimentally unsettled.
Recall that quanta can only be measured and prepared in a single complete context associated
with a  maximal operator (per context).
If the preparation and measurement context coincides,
then ideally the measurement will just reveal the preparation with 100\%
certainty.

Suppose that the two contexts do not coincide.
In this case, for any measurement to take place without a null result,
it could be assumed that the measurement context
may be translated by the measurement apparatus
into the context the quanta have been originally prepared for.
In this scenario, the context translation is carried out by the measurement apparatus alone.
That is, the capability of the measurement apparatus to translate the context
may depend on certain parameters, such as temperature.
Alas, at the moment no models exist which could predict the exact mechanism and
performance of context translation.


\end{document}